\documentclass[twocolumn,prb,showpacs]{revtex4}
\usepackage{graphicx,color}
\usepackage{dcolumn}

\begin{document}
%
\title{
 Ground-state phase diagram of the one-dimensional Hubbard model with an
 alternating potential
}
\author{
 Hiromi Otsuka$^{1}$ and Masaaki Nakamura$^{2}$
}
\address{
 $^1$Department of Physics, Tokyo Metropolitan University, Tokyo
 192-0397 Japan\\
 $^2$Department of Applied Physics, Faculty of Science, Tokyo University
 of Science, Tokyo 162-8601 Japan
} 
\date{\today}
%
\begin{abstract}
 We investigate the ground-state phase diagram of the one-dimensional
 half-filled Hubbard model with an alternating potential---a model for
 the charge-transfer organic materials and the ferroelectric
 perovskites.
 We numerically determine the global phase diagram of this model using
 the level-crossing and the phenomenological renormalization-group
 methods based on the exact diagonalization calculations.
 Our results support the mechanism of the double phase transitions
 between Mott and a band insulators pointed out by Fabrizio, Gogolin,
 and Nersesyan [Phys. Rev. Lett. {\bf 83}, 2014 (1999)]:
 We confirm the existence of the spontaneously dimerized phase as an
 intermediate state.
 Further we provide numerical evidences to check the criticalities on
 the phase boundaries.
 Especially, we perform the finite-size-scaling analysis of the
 excitation gap to show the two-dimensional Ising transition in the
 charge part.
 On the other hand, we confirm that the dimerized phase survives in the
 strong-coupling limit, which is one of the resultants of competition
 between the ionicity and correlation effects.
\end{abstract}
\maketitle
\section{INTRODUCTION}\label{sec_INTRO}

 The electronic and/or magnetic properties of the low-dimensional
 interacting electrons have attracted great interest in researches of
 materials, such as the quasi one-dimensional (1D) organic compounds and
 the two-dimensional (2D) high-$T_{\rm c}$ cuprates, where a variety of 
 generalized Hubbard-type models have been introduced.\cite{Kana63}
 For the 1D case, a concept of the Tomonaga-Luttinger liquid (TLL) has
 been widely accepted and intensively used not only for the descriptions
 on the low-energy and long-distance behaviors of the critical
 systems,\cite{Tomo50,Hald81,Kawa90}
 but also for the prediction of its instabilities to, for instance,
 various types of density-wave phases observed in the
 models.\cite{Voit92}

 The 1D Hubbard model with an alternating potential (also called
 the ionic Hubbard model) is one of the models for the $\pi$-electron
 charge-transfer organic materials, such as TTF-Chloranil,\cite{Naga86}
 and/or the ferroelectric transition metal oxides as
 BaTiO$_3$.\cite{Egam93,Tsuc99}
 It is defined by the Hamiltonian
 \begin{eqnarray}
  H=-t\sum_{j,s}
   \left(
      c^\dagger_{j,s}c^{}_{j+1,s}+{\rm H.c.}
    \right)
   &+&\sum_j U{n_{j,\uparrow}}{n_{j,\downarrow}} \nonumber\\
  &+&\sum_{j}\Delta (-1)^j n_{j}, 
   \label{eq_HAMIL}
 \end{eqnarray}
 where $c_{j,s}$ annihilates an $s$-spin electron ($s=\uparrow$ or
 $\downarrow$) on the $j$th site and the number operator 
 ${n_{j,s}}=c^\dagger_{j,s}c^{}_{j,s}$ and
 $n_j=n_{j,\uparrow}+n_{j,\downarrow}$.
 While $t$ and $U$ terms stand for the electron transfer among sites and
 the Coulomb repulsion on the same site, respectively, the $\Delta$ term
 represents an energy difference between the donor and acceptor
 molecules (or between the cation and oxygen atoms), and it introduces
 ionicity effects into the correlated electron systems (we set $t=1$ in
 the following discussion).

 The understandings on the model have been accumulated in the
 literature, where the theoretical investigations including numerical
 calculations have been performed mainly at the half filling:
 Nagaosa and Takimoto calculated the magnetic and charge-transfer gaps
 as functions of $\Delta$ ($U$ fixed) by using the quantum Monte Carlo
 (QMC) simulation.\cite{Naga86}
 Resta and Sorella, using the exact-diagonalization calculations of
 finite size systems, reported, for instance, the divergence of the
 average dynamical charge.\cite{Rest95} 
 By applying the renormalization-group (RG) method to the bosonized
 Hamiltonian, Tsuchiizu and Suzumura estimated a boundary line between
 the Mott insulator (MI) and a band insulator (BI) phases in the
 weak-coupling regions.\cite{Tsuc99}
 On the other hand, Fabrizio, Gogolin, and Nersesyan (FGN) predicted an
 existence of the ``spontaneously dimerized insulator'' (SDI) phase
 between them.\cite{Fabr99,Fabr00} 
 After their proposal, various numerical calculation methods have been so
 far applied to confirm it:
 Wilkens and Martin performed the QMC simulations to evaluate, e.g., the
 bond order parameter, and reported the transition between the BI and
 SDI phases and stated an absence of MI phase for
 $\Delta>0$.\cite{Wilk01}
 By the combined use of the method of topological transitions (jumps in
 charge and spin Berry phases)\cite{Rest95,Resta98,Alig99,NV02}
 and the method of crossing excitation levels,
 Torio {\it et al.} provided a global ground-state phase diagram, which
 is in accord with the FGN scenario.\cite{Tori01}
 And an existence of the SDI phase for all $U>0$ regions was first
 exhibited there.
 The density matrix renormalization group (DMRG)
 calculations \cite{Taka01,Lou_03,Zhan03,Kamp03} have been performed by
 several groups.
 For instance, Zhang {\it et al.} provided the data on the structure
 factors of relevant order parameters in the weak- and
 intermediate-coupling region, which supports an existence of
 intermediate SDI phase between the BI and MI phases.\cite{Zhan03}
 On one hand, Kampf {\it et al.} estimated the excitation gaps up to
 512-site system and found the boundary of the BI phase while the
 existence of the second boundary was not resolved.\cite{Kamp03}
 Therefore, some controversy as well as points of agreement exists in
 these recent investigations.

 In this paper using the standard numerical techniques, we shall provide
 both the global structure of the ground-state phase diagram and the
 evidences to show the criticalities of the massless spin and charge
 parts.
 For this purpose, it is worthy of noting that the FGN scenario consists
 of two types of instabilities commonly observed in the TLL, i.e., the
 transition with the SU(2)-symmetric Gaussian criticality in the spin
 part, and that with the 2D-Ising criticality in the charge part (see
 Sec.\ \ref{sec_GROUN}).
 Furthermore, these types of phase transitions have been numerically
 treated by the level-crossing (LC) method, and the phenomenological
 renormalization-group (PRG) method.\cite{Room80}
 The LC method has been applied to
 the frustrated XXZ chain,\cite{Okam92,Nomu94}
 and also used in the research of
 higher-$S$ spin chains,\cite{Kita97}
 spin ladders,\cite{Ladder}
 and 1D correlated electron systems.\cite{NakaTJ,NakaEX}
 The advantage of using the LC method is not restricted to its accuracy
 in estimating the continuous phase transition points including the
 Berezinskii-Kosterlitz-Thouless type one; it also provides a means to
 check their criticalities (see Sec.\ \ref{sec_NUMER}).\cite{Nomu94}
 Both of these are important in order to settle the controversy
 mentioned above, and, in fact, the precise estimation of the spin-gap
 transition point of the $S=\frac12$ $J_1$-$J_2$ chain was first given
 by the LC method,\cite{Okam92} while numerical investigations including
 the DMRG work were performed.
 On the other hand, the PRG method is also a reliable numerical approach
 to determine second-order phase transition point, especially for the
 2D-Ising transition where the LC method is not available.
 Analysis based on the PRG method for the 2D-Ising transition is
 successful in the spin systems.\cite{Kita97}
 Furthermore, one of the authors treated the 2D-Ising transition in the
 $S=\frac12$ $J_1$-$J_2$ model under a staggered magnetic field, where
 the critical phenomena in the vicinity of the phase boundary line were
 argued.\cite{Otsu02} 
 Therefore, based on these recent developments, we shall perform the
 numerical calculations; especially, to our knowledge, this is the first
 time that the PRG method successfully applied to the 2D-Ising
 transition observed in one part of the two-component systems like the
 interacting electrons.


 The organization of this paper is as follows. 
 In Sec.\ \ref{sec_GROUN}, 
 we shall briefly refer to the effective theory based on the bosonized
 Hamiltonian and order parameters of expected density-wave phases, and
 mention the FGN scenario.
 In Sec.\ \ref{sec_NUMER}, 
 we explain procedures of the numerical calculation to determine
 transition lines, where connections between the methods and
 instabilities of the TLL systems will be explained.
 After that, we shall give a ground-state phase diagram in whole
 parameter region.
 Furthermore, to confirm the criticalities and to serve a reliability of
 our calculations, we check the consistency of excitation levels in
 finite-size systems.
 A finite-size scaling analysis of the charge excitation gaps is also
 performed in the vicinity of the phase boundary line.
 Section\ \ref{sec_DISCU} is devoted to discussions and summary of the
 present investigation.
 A short comment on the Berry phase method
 \cite{Rest95,Resta98,Alig99,NV02,Tori01}
 will also be given there.
 We will provide the comparison with that method, which is helpful to
 exhibit a reliability of our approach as well as the results.

\section{GROUND STATES AND PHASE TRANSITIONS}\label{sec_GROUN}

 The bosonization method provides an efficient way to describe
 low-energy properties of the 1D quantum systems:\cite{Gogo98}
 Linearizing the $\cos$-band at two Fermi points $\pm k_{\rm F}=\pm\pi
 n/2a$ [an electron density $n:=N/L=1$ and a number of sites (electrons)
 $L$ ($N$)], and according to standard procedure, the effective
 Hamiltonian \cite{Tsuc99,Fabr99,Fabr00} is given as $H\rightarrow{\cal
 H}={\cal H}_\rho+{\cal H}_\sigma+{\cal H}_2$
 with
 \begin{eqnarray}
  {\cal H}_\nu
   &=&
   \int {d}x~\frac{v_\nu}{2\pi}
   \left[
    {       K_\nu} \left(\partial_x \theta_\nu\right)^2 + 
    {1\over K_\nu} \left(\partial_x   \phi_\nu\right)^2  
  \right]\nonumber\\ 
  &+&
   \int {d}x~\frac{2g_\nu}{(2\pi\alpha)^2}~
   {\cos\sqrt8\phi_\nu},~~~(\nu=\rho, \sigma),
   \label{eq_Hubba}\\ 
  {\cal H}_2
   &=&
   \int {d}x~\frac{-2\Delta}{\pi\alpha}~
   {\sin\sqrt2\phi_\rho}~{\cos\sqrt2\phi_\sigma}.
   \label{eq_alter}
 \end{eqnarray}
 The operator $\theta_\nu$ is the dual field of $\phi_\nu$
 satisfying the commutation relation
 $
 \left[\phi_\nu(x),\partial_y\theta_{\nu'}(y)/\pi\right]=
 {i}\delta(x-y)\delta_{\nu,\nu'}.
 $
 $K_\nu$ and $v_\nu$ are the Gaussian coupling and the velocity of
 elementary excitations. 
 Coupling constants $g_\rho~(<0)$ and $g_\sigma$ stand for the 4$k_{\rm
 F}$-Umklapp scattering and the backward scattering bare amplitudes,
 respectively, and ${\cal H}_2$ expresses a coupling between the spin
 and charge degrees of freedom.
 In Table\ \ref{TAB_I}, we summarize the order parameters for the
 relevant $2k_{\rm F}$ density-wave phases, i.e.,
 the charge-density-wave (CDW),  
 bond charge-density-wave (BCDW), and
 spin-density-wave (SDW) phases, where 
 the electron's spin and the bond charge are given as
 $
 {\bf S}_j=\sum_{s,s'}
 c^\dagger_{j,s}[\frac12\mbox{\boldmath$\sigma$}]^{}_{s,s'}c^{}_{j,s'}
 $
 and
 $\overline{n}_j=\sum_s (c^\dagger_{j,s} c^{}_{j+1,s}+{\rm H.c.})$,
 respectively (\mbox{\boldmath$\sigma$} are the Pauli matrices). 
 Their bosonized expressions are given in the second column. 
 In the third column, we give the locking points of phase fields.
 As discussed in Ref.\ \onlinecite{Fabr99}, there are two locking points
 of $\phi_\rho$, i.e.,
 $\langle\sqrt8\phi_\rho\rangle=\pm\phi_0$ in the BCDW state.
 The phase $\phi_0$, a function of $U$ and $\Delta$, continuously varies
 from 0 to $\pi$.

 Let us see the system with increasing $\Delta$ for fixed $U$.
 At $\Delta=0$, the ground state is in the MI phase with the most
 divergent SDW fluctuation (the third row of Table\ \ref{TAB_I}).
 According to the arguments,\cite{Tsuc99,Naga86} the MI phase may
 survive for $U\gg2\Delta$. 
 For $2\Delta\gg U$, ${\cal H}_2$ becomes relevant, and leads to the BI
 phase with the long-range CDW order without degeneracy (the first row).
 For this issue, FGN argued that under the uniform charge distribution a
 renormalization effect of ${\cal H}_2$ to $g_\sigma$ brings about the
 spin-gap transition in the spin part at a certain value of
 $\Delta_\sigma(U)$, which is described by the sine-Gordon (SG) theory.
 This is qualitatively in accord with the perturbation calculation in
 the strong-coupling region,\cite{Naga86} and leads to the SDI phase
 with the long-range BCDW order (the second row).
 Further with the increase of $\Delta$, a transition in the charge part
 occurs on a separatrix $\Delta_\rho(U)$ between two different types of
 charge-gap states.
 This line corresponds to the massless RG flow connecting the Gaussian
 (the central charge $c=1$) and the 2D-Ising ($c=\frac12$) fixed
 points,\cite{Zamo86}
 and its description is given by the double-frequency sine-Gordon (DSG)
 theory.\cite{Delf98}
 Our main task is thus to estimate $\Delta_\nu(U)$ for $U>0$ and to
 check the criticalities based on their prediction.
\begin{table}
 \caption{
 The order parameters. 
 The bosonized forms and the locking points of phase variables
 $
 (\protect\langle\protect\sqrt8\protect\phi_{\protect\rho  }\protect\rangle,
  \protect\langle\protect\sqrt8\protect\phi_{\protect\sigma}\protect\rangle)$
 are given in the second and third columns. 
 $\phi_0$ is a function of $U$ and $\Delta$, and $\ast$ denotes a phase
 not to be locked.
 }
 \begin{tabular}{crccc}\hline\hline
  &
  Order parameters
  &
  Bosonized forms
  &
  Locking points
  &
  \\
  \tableline
  & ${\cal O}_{\rm CDW}=             (-1)^jn_j$                      &
  $2\sin\sqrt2\phi_\rho \cos\sqrt2\phi_\sigma$ & $(    \pi   ,   0)$ &\\
  & ${\cal O}_{\rm BCDW}=  (-1)^j\overline{n}_j$    &
  $2\cos\sqrt2\phi_\rho \cos\sqrt2\phi_\sigma$ & $(\pm \phi_0,   0)$ &\\
  & ${\cal O}^{\parallel}_{\rm SDW}= (-1)^jS^{z}_j$           &
  $2\cos\sqrt2\phi_\rho \sin\sqrt2\phi_\sigma$ & $(         0,\ast)$ &\\
  \hline\hline
 \end{tabular}
 \label{TAB_I}
\end{table}

\section{NUMERICAL METHODS AND CALCULATION RESULTS}\label{sec_NUMER}

 Low-lying excitations observed in the finite-size systems are expected
 to serve for the determinations of transition points.
 Here, we take a look at the following operators with lower scaling
 dimensions:
 \begin{eqnarray}
  {\cal O}_{\nu,1}&=&\sqrt2\cos\sqrt2\phi_\nu,
   \label{eq_COS}     \\ 
  {\cal O}_{\nu,2}&=&\sqrt2\sin\sqrt2\phi_\nu,
   \label{eq_SIN}     \\ 
  {\cal O}_{\nu,3}&=&{\rm exp(\pm i}\sqrt2\theta_\nu).
   \label{eq_EXP}
 \end{eqnarray}
 According to the finite-size-scaling argument based on the conformal
 field theory, corresponding energy levels for these operators $\Delta
 E_{\nu,i}$ (taking the ground-state energy $E_0$ as zero) are expressed
 by the use of their scaling dimensions $x_{\nu,i}$:\cite{Card84}
 \begin{equation}
  \Delta E_{\nu,i}\simeq \frac{2\pi v_\nu}{L}x_{\nu,i}.
   \label{eq_SCALING}
 \end{equation}
 Then these excitations can be extracted under the antiperiodic boundary
 condition with respect to the ground state due to the selection rule
 of the quantum numbers.\cite{NakaTJ,NakaEX}
 In the numerical calculations using the Lanczos
 algorithm we can identify $\Delta E_{\nu,i}$
 according to
 the discrete symmetries of the wave functions, e.g.,
 translation ($c_{j,s}\to c_{j+2,s}$),
 charge conjugation [$c_{j,s}\to (-1)^jc_{j+1,s}^{\dag}$],
 spin reverse ($c_{j,s}\to c_{j,-s}$), and
 space inversion ($c_{j,s}\to c_{L-j,s}$).
 Here note that, except for the spin-reversal operation, definitions of
 these transformations are different from those of the uniform systems,
 such as the extended Hubbard model.\cite{NakaEX}

\begin{figure}[t]
 \begin{center}
  \includegraphics[width=3.2in]{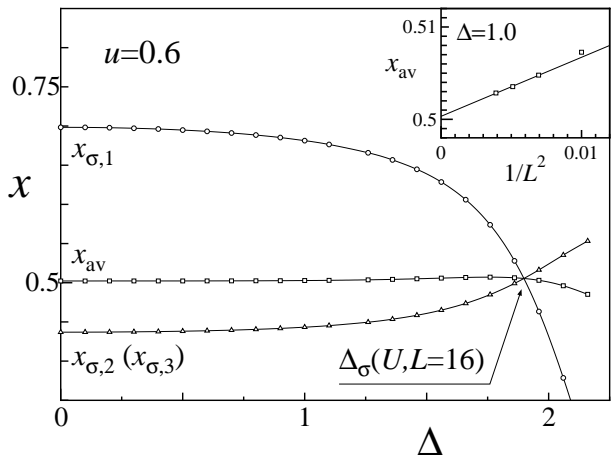}
 \end{center}
 \caption{
 The $\Delta$ dependence of $x_{\sigma,i}$ at $u=0.6$ for the 16-site
 system [$u=U/(U+4)$].
 The spin-gap transition point $\Delta_\sigma(U,L)$ is estimated from
 the level crossing between the singlet (circles) and triplet
 (triangles) spin excitations.
 The squares plot $x_{\rm av}=(x_{\sigma,1}+3 x_{\sigma,2})/4$,
 and the inset shows the $L$ dependence of $x_{\rm av}$ at $\Delta=1.0$,
 where a least-square-fitting line to the data of $L=12$-$16$ is given.
 }
 \label{FIG1}
\end{figure}

 First, we treat the spin-gap transition in the spin part following
 Refs.\ \onlinecite{Okam92,NakaTJ}, and \onlinecite{NakaEX}.
 In the SDW phase, due to the marginal coupling in the SU(2)-symmetric
 spin part, the singlet ($x_{\sigma,1}$) and triplet
 ($x_{\sigma,2}=x_{\sigma,3}$) excitations split as
 $x_{\sigma,1}>x_{\sigma,2}=x_{\sigma,3}$ satisfying a universal
 relation
 \begin{equation}
 \frac{x_{\sigma,1}+3 x_{\sigma,2}}{4}=\frac12. 
  \label{eq_SPN13}
 \end{equation}
 Then, the degeneracy condition
 \begin{equation}
  x_{\sigma,1}=x_{\sigma,2}=x_{\sigma,3}
   \label{eq_SPNCR}
 \end{equation}
 stands for the vanishing of the coupling, and provides a good estimation
 of the spin-gap transition point.
 Note that Torio {\it et al.} used the crossing of these excitation
 levels for the determination of the MI-SDI transition,\cite{Tori01}
 while the consistency check of the levels to confirm the universality
 of transition is still absent.
 Figure\ \ref{FIG1} shows an example of the $\Delta$ dependences of
 $x_{\sigma,i}$ for the 16-site system at $u=0.6$ [here we introduce the
 reduced Coulomb interaction parameter $u=U/(U+4)$].
 For this plot, we estimated the spin-wave velocity $v_\sigma$ from a
 triplet excitation with the wave number $4\pi/L$ as
 $v_{\sigma}=\lim_{L\to\infty}\Delta E(S=1,k=4\pi/L)/(2\pi/L)$
 and normalized the excitation gaps $\Delta E_{\sigma,i}$ according to
 Eq.\ (\ref{eq_SCALING}).
 The singlet (triplet) level corresponding to the operator ${\cal
 O}_{\sigma,1}$ [${\cal O}_{\sigma,2}~({\cal O}_{\sigma,3})$] is denoted
 by circles (triangles) with a fitting curve. 
 Their behaviors reflect the TLL properties:
 For instance, the amplitude of the level splitting decreases with the
 increase of $\Delta$ due to its renormalization effect, and eventually
 the level crossing occurs at $\Delta_\sigma(U,L)$.
 More precisely, in order to confirm the universality, we plot the
 averaged scaling dimension $x_{\rm av}$, i.e., the left-hand side of
 Eq.\ (\ref{eq_SPN13}) in Fig.\ \ref{FIG1} (squares).
 We also exhibit the $L$ dependence of $x_{\rm av}$ at $\Delta=1.0$ as
 an example (see the inset).
 The result shows that the condition imposed on $x_{\sigma,i}$ is
 accurately satisfied for $\Delta\le\Delta_\sigma(U,L)$; in particular, 
 the extrapolated value of $x_{\rm av}$ is almost $\frac12$.
 Consequently, the level crossing at which Eq.\ (\ref{eq_SPNCR}) is
 satisfied can be regarded as an indication of the spin-gap transition
 in the spin part of the Hamiltonian (\ref{eq_HAMIL}).
 On the other hand, the spin part is dimerized for
 $\Delta>\Delta_\sigma(U,L)$.

\begin{figure}[t]
 \begin{center}
  \includegraphics[width=3.2in]{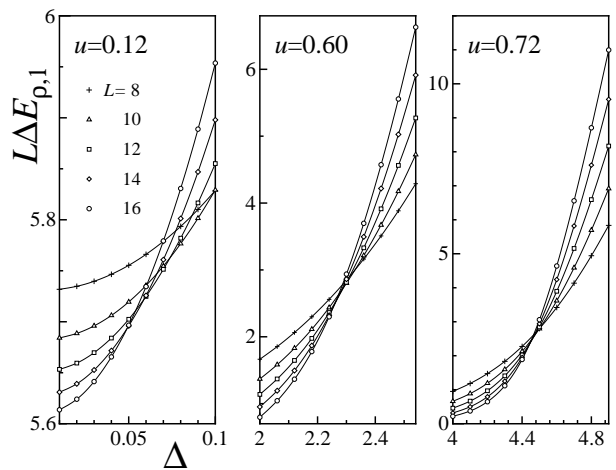}
 \end{center}
 \caption{
 The $L$ and $\Delta$ dependences of the scaled gap $L\Delta E_{\rho,1}$.
 From left to right, $u=$0.12, 0.60 and 0.72, respectively. 
 The correspondence between marks and system sizes is given in the figure. 
 Crossing points give the $L$-dependent transition points $\Delta_\rho(U,L+1)$.
 }
  \label{FIG2}
\end{figure}

 Next, we discuss the 2D-Ising transition in the charge part. 
 Recently, we have treated the crossover behavior into the 2D-Ising
 criticality in the study of the frustrated quantum spin
 chain,\cite{Otsu02} so we shall here employ the same approach to
 determine $\Delta_\rho(U)$.
 Since there are two critical fixed points connected by the RG flow, 
 a relationship between lower-energy excitations on these fixed points
 is quite important. 
 For this, the so-called ultraviolet-infrared (UV-IR) operator
 correspondence provides significant informations:\cite{Fabr00,Bajn01}
 Along the RG flow, the operators on the Gaussian fixed point (UV) are
 transmuted to those on the 2D-Ising fixed point (IR) as
 \begin{equation}
  {\cal O}_{\rho,1} \to \mu,~~~
  {\cal O}_{\rho,2} \to I+\epsilon,
   \label{eq_UVIRC}
 \end{equation}
 where $\mu$ is the disorder field (Z$_2$ odd), and $\epsilon$ is the
 energy density operator (Z$_2$ even) with scaling dimensions
 $x_\mu=\frac18$ and $x_\epsilon=1$, respectively.
 Furthermore, since a deviation from the transition point
 $\Delta-\Delta_\rho(U)$,
 which is the coupling constant of the ${\cal O}_{\rho,2}$ term
 in the DSG Hamiltonian,\cite{Fabr99}
 plays a role of the thermal scaling variable, anomalous behaviors in
 the vicinity of $\Delta_\rho(U)$ are to be related to the divergent
 correlation length of the form $\xi\propto
 [\Delta-\Delta_\rho(U)]^{-\nu}$ with the exponent
 $1/\nu=2-x_\epsilon=1$.
 On one hand, the excitation $\mu$ corresponding to ${\cal O}_{\rho,1}$
 provides a lower-energy level, so we shall focus our attention on it.
 In order to determine the transition point, we shall numerically solve
 the following PRG equation for a given value of $U$
 with respect to $\Delta$:\cite{Room80,Otsu02}
 \begin{equation}
  (L+2)\Delta E_{\rho,1}(U,\Delta,L+2)=
   L   \Delta E_{\rho,1}(U,\Delta,L  ).
   \label{eq_PRGEQ}
 \end{equation}
 Since this is satisfied by the gap
 $\Delta E_{\rho,1}(U,\Delta,L)\propto 1/L$,
 the obtained value can be regarded as the $L$-dependent transition
 point, say $\Delta_\rho(U,L+1)$.
 We plot $L$ and $\Delta$ dependences of the scaled gap $L\Delta
 E_{\rho,1}(U,\Delta,L)$ in Fig.\ \ref{FIG2}, and find that the
 size dependence of the crossing point is small for large values
 of $U$, but it is visible in the weak coupling case. 

\begin{figure}[t]
 \begin{center}
  \includegraphics[width=3.2in]{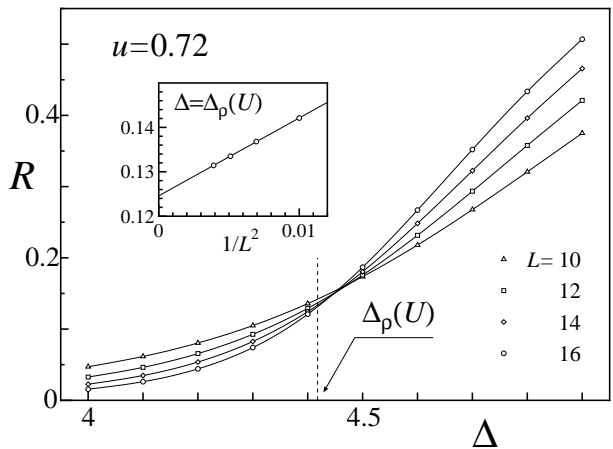}
 \end{center}
 \caption{
 The $\Delta$ dependence of the charge-excitation-gap ratio
 $R={\Delta E_{\rho,1}(U,\Delta,L)}/{\Delta E_{\rho,2}(U,\Delta,L)}$
 for $L=10$-$16$ at $u=0.72$.
 The arrow shows the transition point $\Delta_\rho(U)$.
 The inset plots the $L$ dependence of $R$ at $\Delta_\rho(U)$ with a
 least-square-fitting line.
 }
 \label{FIG3}
\end{figure}

 While the results in the thermodynamic limit will be given in the last
 part of this section, we shall check first the criticality on and in
 the vicinity of the phase boundary using the extrapolated data
 $\Delta_\rho(U)$.
 For this aim, an evaluation of the central charge $c$ through the size
 dependence of the ground-state energy provides a straightforward
 way.\cite{Blot86}
 However, as exhibited in the following, the critical line in the charge
 part is close to the spin-gap transition line, so that influences from
 the spin part with the small dimer gap prohibit a reliable estimation
 of $c$ from the data of the finite-size systems.
 Alternatively, we shall evaluate a ratio of the charge-excitation gaps
 $\Delta E_{\rho,1}(U,\Delta,L)$ and
 $\Delta E_{\rho,2}(U,\Delta,L)$
 on the phase boundary to check the UV-IR operator correspondence. 
 According to Eqs.\ (\ref{eq_SCALING}) and (\ref{eq_UVIRC}), it is
 expressed by the scaling dimensions of operators $\epsilon$ and $\mu$
 as
 \begin{equation}
  R=
  \frac
   {\Delta E_{\rho,1}(U,\Delta_\rho(U),L)}
   {\Delta E_{\rho,2}(U,\Delta_\rho(U),L)}
   \to 
   \frac
   {x_\mu}
   {x_\epsilon}=\frac18
 \end{equation}
 for large $L$.
 Figure\ \ref{FIG3} plots the $\Delta$ dependence of $R$ for $L=10$-$16$
 ($u=0.72$).
 The transition point in the thermodynamic limit is denoted by the arrow
 near the $x$ axis.
 While the ratio exhibits a subtle $\Delta$ dependence around the point,
 we interpolate these data, and estimate the $L$ dependence of $R$ at
 $\Delta_\rho(U)$, which is given with a least-square-fitting line in
 the inset.
 The plot shows that the extrapolated value is fairly close to
 $\frac18$.
 Therefore we conclude that the boundary line $\Delta_\rho(U)$ belongs
 to the 2D-Ising universality class.

\begin{figure}[t]
 \begin{center}
  \includegraphics[width=3.2in]{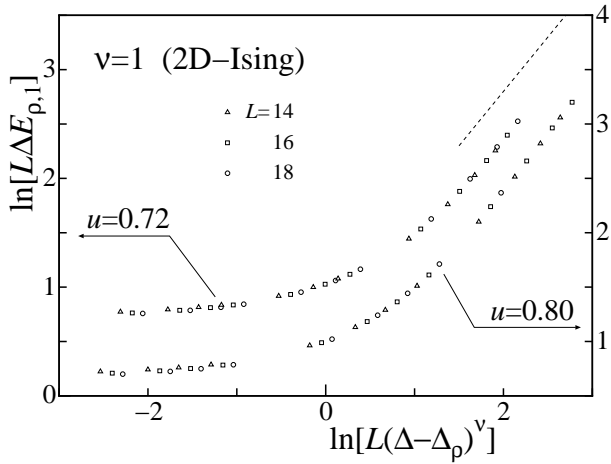}
 \end{center}
 \caption{
 The finite-size-scaling plots of the charge-excitation gap
 $\Delta E_{\rho,1}$ for systems of $L=14$-$18$ at $u=0.72$ and 0.80.   
 We use the 2D-Ising critical exponent $\nu=1$. 
 A dotted line (the slope 1) is given for the guide to eye.
 }
 \label{FIG4}
\end{figure}

 Furthermore, we shall investigate the critical behavior:\cite{Otsu02}
 According to the finite-size-scaling argument,
 we analyze the charge-excitation gap by using the following
 one-parameter scaling form:
 \begin{equation}
  \Delta E_{\rho,1}(U,\Delta,L)=L^{-1}\Psi(L[\Delta-\Delta_\rho(U)]^\nu).
   \label{eq-FSS}
 \end{equation}
 Since $\Delta E_{\rho,1}\propto 1/\xi$ in the thermodynamic limit
 ($L/\xi\to\infty$), the scaling function is expected to asymptotically
 behave as $\Psi(x)\propto x$ for large $x$.
 On the other hand, the gap $\Delta E_{\rho,1}\propto 1/L$ on the
 critical point ($L/\xi\to0$) so that $\Psi(x)\simeq {\rm const}$ for
 $x\to0$.\cite{Barb83}
 Figure\ \ref{FIG4} plots Eq.\ (\ref{eq-FSS}) using the exponent of
 the 2D-Ising model $\nu=1$. 
 Although due to the smallness of $L$ a scattering of the scaled data is
 visible especially near the transition point, the data of different
 system sizes are collapsed on the single curve, and its asymptotic
 behaviors agree with the expected ones.
 Therefore, we can check that, in the transition of the charge part, the
 deviation $\Delta-\Delta_\rho(U)$ plays a role of the thermal scaling
 variable on the 2D-Ising fixed point.
 Here, note that in the strong-coupling region the energy scale of the
 crossover behavior may be large enough to be detected even in the
 small-size systems. 
 However, the finite-size-scaling nature may become obscure in the weak
 and intermediate couplings.

\begin{figure}[t]
 \begin{center}
  \includegraphics[width=3.2in]{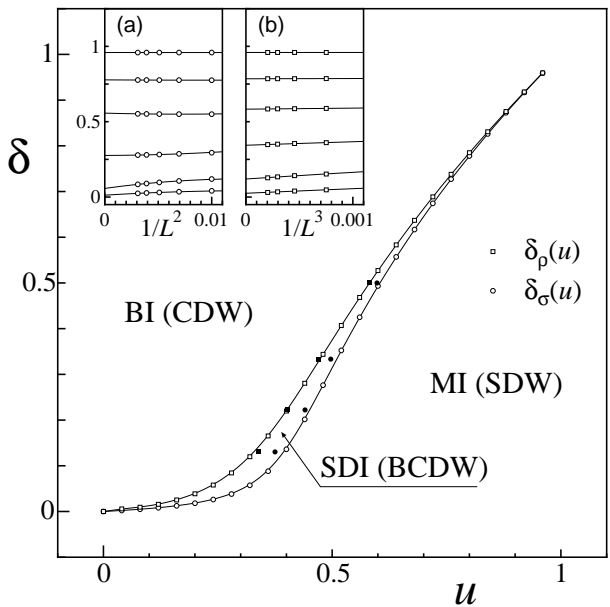}
 \end{center}
 \caption{
 The ground-state phase diagram of the 1D Hubbard model with the
 alternating potential.
 The open circles (squares) with a fitting curve show the spin-gap
 (2D-Ising) transition line in the spin (charge) part. 
 The stable regions of
 the MI, SDI, and BI phases are given in the 2D parameter space $(u,\delta)$
 [$u=U/(U+4)$ and $\delta=\Delta/(\Delta+2)$].
 Insets (a) and (b) show the extrapolations of the $L$-dependent
 transition points in the spin and the charge parts, respectively.
 For comparison, we also plot the DMRG calculation results given in
 Ref.\ \onlinecite{Zhan03} by using the filled squares ($U_{c1}$ in
 their notation) and the filled circles ($U_{c2}$). 
 }
  \label{FIG5}
\end{figure}

\begin{figure}[t]
 \begin{center}
  \includegraphics[width=3.2in]{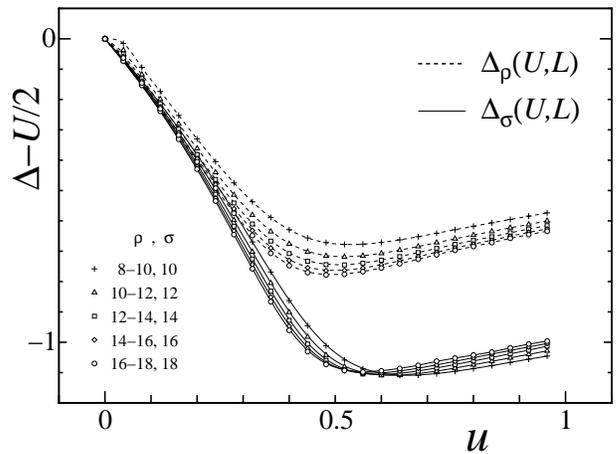}
 \end{center}
 \caption{
 The deviations of boundaries from the $2\Delta=U$ line,
 $\Delta_\nu(U,L)-U/2$.
 We use $u=U/(U+4)$ as the $x$ axis.
 The correspondence between marks and system sizes is given in the figure. 
 Marks with solid (dotted) curves exhibit the deviations in the spin
 (charge) part.
 }
  \label{FIG6}
\end{figure}

 Lastly, we present the ground-state phase diagram. 
 In order to determine it, the extrapolations of $\Delta_\nu(U,L)$ to
 the thermodynamic limit are carried out.
 For the spin part, it should be noted that Torio {\it et al.} evaluated
 the spin-gap transition line from the level crossing Eq.\
 (\ref{eq_SPNCR}),\cite{Tori01} so here we perform the same calculations
 in order to complete the ground-state phase diagram.
 We employ the formula: $\Delta_{\sigma}(U,L)=\Delta_{\sigma}(U)+a
 L^{-2} +b L^{-4}$, where $\Delta_{\sigma}(U)$, $a$ and $b$ are
 determined according to the least-square-fitting condition. 
 Then, we extrapolated the data of $L=12$-$18$ as shown in Fig.\
 \ref{FIG5}(a), where from bottom to top the data with fitting 
 curves are given in the increasing order of $U$.
 Consequently, the spin-gap transition line $\Delta_{\sigma}(U)$
 (open circles with a fitting curve) is given in Fig.\ \ref{FIG5}, where
 the reduced alternating potential parameter $\delta=\Delta/(\Delta+2)$
 is used as the $y$ axis.
 On the other hand, for the extrapolation of $\Delta_\rho(U,L)$, we
 assume the following formula:\cite{Itzy89}
 $\Delta_{\rho}(U,L)=\Delta_{\rho}(U)+a L^{-3}$, and extrapolate the
 data of $L=10$-18 as shown in Fig.\ \ref{FIG5}(b).
 Consequently, Fig.\ \ref{FIG5} shows that the critical line in the
 charge part (open squares with a fitting curve) does not coincide with
 the spin-gap transition line, i.e.,
 $\Delta_{\sigma}(U)<\Delta_{\rho}(U)$, and that the 2D parameter space
 $\{(u,\delta)~|~0\le u,~\delta\le1\}$ is separated into the MI, BI, and
 SDI phases with SDW, CDW, and BCDW, respectively.
 Since the Hubbard gap provides a principal energy scale and a shape of
 the boundary is roughly determined so that the magnitude of the band
 gap becomes comparable to the scale, the $U$ dependence of the
 boundaries is expected to be weak in the small-$U$
 region,\cite{Tsuc99,Fabr99}
 which is in agreement with our observation.
 On the other hand, in order to clarify the behaviors in the large-$U$
 region, we plot a magnification of the phase diagram around the
 $2\Delta=U$ line in Fig.\ \ref{FIG6}.
 This shows that in the limit of $U\to\infty$ the boundaries do not
 merge to the line: More precisely,
 for $U=96$ we obtain 
 $\Delta_{\rho}-U/2 \simeq -0.65$ and 
 $\Delta_{\sigma}-U/2 \simeq -0.97$, respectively.
 In Ref.\ \onlinecite{Tori01}, adding to the spin part
 ($2\Delta_{\sigma}-U \simeq -1.91$ for $U$, $V\gg1$), they also
 reported $2\Delta_{\rho}-U \simeq -1.33$, which is close to our
 estimation.
 Consequently, we confirm that the intermediate SDI phase may
 survive in the large-$U$ limit, which is one of the nontrivial
 behaviors and is contrasted to the naive argument.

 Here we shall perform a comparison with the previous DMRG results. 
 As mentioned in Sec.\ \ref{sec_INTRO}, while the DMRG calculations
 performed by several groups seem not to reach an agreement with respect
 to an existence of the SDI phase, it may be informative to provide a
 comparison with our result.
 Zhang {\it et al.} determined two-types of phase transition points
 $U_{c1}$ and $U_{c2}$ based on the structure factor of the BCDW order
 parameter;\cite{Zhan03}
 we plot their results in Fig.\ \ref{FIG5} by using the filled squares
 and filled circles, respectively.
 This shows that their estimations of $U_{c1}$ agree well with our data
 $\Delta_\rho(U)$, although those of $U_{c2}$ considerably deviate from
 $\Delta_\sigma(U)$.
 Since the phase transition at $\Delta_\sigma(U)$ is the spin-gap
 transition, the logarithmic corrections to the power-law behaviors
 as well as the exponentially small magnitude of the spin gap generally
 make it difficult to determine the transition point. 
 On one hand, as explained in the above, the LC method used here
 overcomes these difficulties in the determination of the transition
 points $\Delta_\sigma(U)$.


\section{DISCUSSION and SUMMARY}\label{sec_DISCU}

 For the understanding of the phase diagram in the large-$U$ limit,
 let us see the perturbative treatment of Hamiltonian\ (\ref{eq_HAMIL})
 under the condition of $U-2\Delta\gg1$.
 $\Delta_\sigma(U)$ may be related to the spin-gap transition point in
 the $S=\frac12$ $J_1$-$J_2$ model.\cite{Fabr99,Kamp03}
 Therefore, using its numerical value\cite{Okam92}
 and perturbative expressions on $J_1$ and $J_2$,\cite{Naga86}
 we can approximately estimate $\Delta_\sigma(U)$
 as a solution of the equation $J_2/J_1\simeq X/(1-4X)\simeq0.2411$,
 where $X=(1+4x^2-x^4)/U^2(1-x^2)^2$ and $x=2\Delta/U$.
 Then, we find a solution [$\Delta'_\sigma(U)$] to give a value
 $\Delta'_\sigma(U)-U/2\simeq-1.427$ in the limit.
 While, due to the lack of effects from the higher-order processes in
 the kinetic energy term, the approximate value deviates from the
 numerical estimation, this exhibits the following, i.e., the
 perturbative expansion becomes singular on the $2\Delta=U$ line so that
 the phase boundary deviates from the line.
 This singularity also exists in the perturbative calculations of the
 SDW and CDW state energies ($E_{\rm SDW}$ and $E_{\rm CDW}$). 
 And then the direct transition line between these phases cannot be
 determined from the equation $E_{\rm SDW}=E_{\rm CDW}$, which is highly
 contrasted to the case of the extended Hubbard model (EHM) including
 the nearest-neighbor Coulomb interaction $\sum_j V
 n_jn_{j+1}$.\cite{Dong94}
 Since the spin-charge coupling term with the dimerized spin part
 generates one of the relevant forces, $\Delta_\rho(U)$ should be
 affected by that of the spin part.
 Besides the present model, it is known that EHM possesses the coupling
 term  
 $V\cos\sqrt8\phi_\rho\cos\sqrt8\phi_\sigma$ in its bosonized
 form,\cite{Voit92}
 and that the BCDW state with the locking points 
 $\langle\sqrt8\phi_{\rho,\sigma}\rangle=0$ 
 is stabilized around the $2V=U$ line in the weak- and
 intermediate-coupling region.\cite{NakaEX}
 The corrections to $g_\nu$ from higher-energy states stabilize
 it,\cite{Tsuc01}
 but the coupling term forces the boundaries to merge into
 the single first-order phase transition line between the SDW and CDW
 states in the strong-coupling region because it raises the BCDW state
 energy.
 However, in the present BCDW state, the locking point $\phi_0$ in
 Table\ \ref{TAB_I} may take a value so as not to bring about a large
 energy cost due to the coupling term Eq.\ (\ref{eq_alter}).
 Therefore, the existence of the SDI phase is not prohibited even in the
 strong-coupling limit in contrast to the EHM case.
 Of course, these arguments are qualitative and intuitive ones, so an
 effective theory in this limit is required for the precise description
 on the limiting behaviors.

\begin{figure}[t]
 \begin{center}
  \includegraphics[width=3.2in]{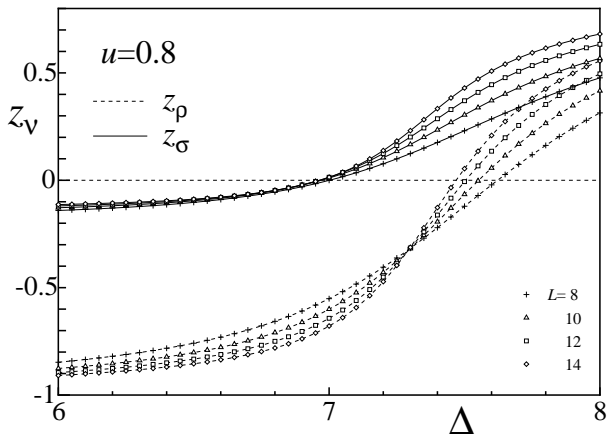}
 \end{center}
 \caption{
 Behavior of the ground-state expectation value of the twist operator
 $z_{\nu}$ ($\nu=\rho,\sigma$) near the $2\Delta=U$ line.
 The correspondence between marks and system sizes are given in the
 figure. 
 }
  \label{FIG7}
\end{figure}
\begin{figure}[t]
 \begin{center}
  \includegraphics[width=3.2in]{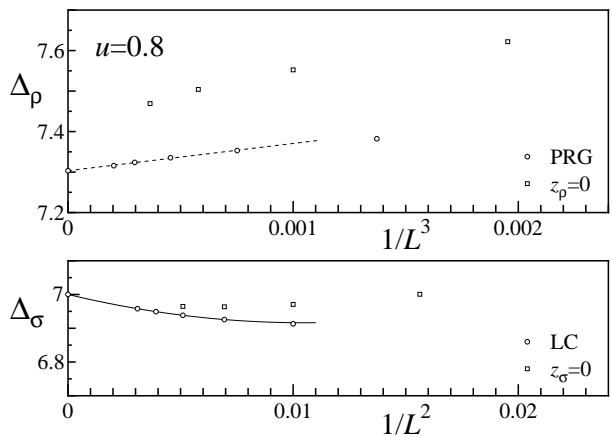}
 \end{center}
 \caption{
 Comparisons of the system-size dependences of the transition points
 obtained by the LC and PRG methods vs by the condition $z_{\nu}=0$.
 The fitting curves show the extrapolations of data to the thermodynamic
 limit.
 }
  \label{FIG8}
\end{figure}

 Finally, we comment on the Berry phase
 method.\cite{Rest95,Resta98,Alig99,NV02,Tori01}
 The Berry phases for the charge and the spin parts $\gamma_\nu$ are
 related to the ground-state expectation values of the twist operators
 as $\gamma_\nu={\rm Im}\log z_\nu$ where
 \begin{equation}
  z_{\rho}=\langle U_{\uparrow}U_{\downarrow}\rangle,\quad
  z_{\sigma}=\langle U_{\uparrow}U_{\downarrow}^{-1}\rangle,
 \end{equation}
 and $U_{s}=\exp[(2\pi i/L)\sum_{j=1}^L j n_{j,s}]$.\cite{Resta98}
 Since $z_\nu$ is real at the half filling with zero-magnetic field,
 $\gamma_\nu$ (=0 or $\pi$) indicates the sign of $z_\nu$.
 On one hand, $z_\nu$ can be related to the bosonic field as
 $z_{\rho,\sigma}\propto\mp\langle\cos\sqrt{8}\phi_{\rho,\sigma}\rangle$,
 so that it includes the information of the locking points given in
 Table\ \ref{TAB_I}.\cite{NV02}
 In Fig.\ \ref{FIG7} we show behaviors of $z_\nu$ near the $2\Delta=U$
 line for $U=16$ and find that with the increase of $\Delta$ both of
 these increase and change their sign.
 As shown in the lower panel of Fig.\ \ref{FIG8}, the condition
 $z_\sigma=0$ gives a close value to the result of the LC method, so
 it may provide a proper estimation of the spin-gap transition point
 $\Delta_\sigma$.\cite{Tori01,NV02}
 On the other hand, the zero point of $z_\rho$ exhibits a deviation from
 the PRG result (see the upper panel of Fig.\ \ref{FIG8}).
 Since $\phi_0$ continuously varies with $\Delta$, $z_\rho$ can take a
 finite value on the 2D-Ising transition point in the thermodynamic
 limit, which is highly contrasted to $z_\sigma$ on the spin-gap
 transition point.
 In fact, the size-dependent zero points are seemingly extrapolated to a
 value different from our PRG estimation, so that the condition $z_\rho=0$
 does not specify the transition point.
 On the other hand, we also find in Fig.\ \ref{FIG7} that there is a
 point $\Delta\simeq 7.3$ at which $z_\rho$ is almost independent of
 $L$.
 This crossing point is expected to be a good estimator for the 2D-Ising
 transition point in the charge part $\Delta_\rho$  because this is
 quite close to the PRG result even for small $L$.
 However, a theoretical explanation of this possibility is still open.  


 To summarize, we have investigated the ground-state phase diagram of
 the one-dimensional half-filled Hubbard model with the alternating
 potential, especially in order to verify the scenario given by
 Fabrizio, Gogolin, and Nersesyan, we have numerically treated the phase
 transitions observed in the spin and charge parts:
 We calculated the spin-gap transition points $\Delta_\sigma$ in the
 spin part by the level-crossing method (see also the argument for the
 spin-gap transition in Ref.\ \onlinecite{Tori01}) and the
 two-dimensional Ising transition points $\Delta_\rho$ in the charge
 part by the phenomenological renormalization-group method.
 We confirmed that, adding to the Mott and band insulators, the
 ``spontaneously dimerized insulator'' accompanied by the
 long-range-ordered $2k_{\rm F}$ bond charge-density wave is stabilized
 as the intermediate phase for all $U>0$.
 Then we checked the SU(2)-symmetric Gaussian (2D-Ising) criticality of
 the spin (charge) part by treating the low-lying excitation levels in
 the finite-size systems, and, simultaneously, we performed the
 finite-size-scaling analysis of the charge-excitation gap to clarify
 the critical phenomena around $\Delta_\rho$.
 The comparison with the relating work was performed to check the
 reliability of our numerical results and to exhibit the efficiency of
 our approach.


 After submission of this paper, we became aware of the work
 investigating the ground-state phase diagram and the universality of
 the transition in the charge part by the use of finite-size-scaling 
 analysis of the DMRG calculation data.\cite{Manm03}
 They have found two transition points and succeeded to obtain $\nu=1$
 in agreement with our conclusion, while the estimated exponent for the
 susceptibility of the BCDW order parameter shows a deviation from the
 theoretical value $\eta_1=1/4$,
 e.g., $\eta_1\simeq0.45$ at the point on the BI-SDI
 phase boundary $\Delta=10$ and $U_{c1}=21.385$ (in their notation).
 In this paper we have treated the elementary excitations in the TLL
 system specified by the discrete symmetries of the lattice Hamiltonian
 with the twisted boundary condition, whereas they have measured the BCDW
 order parameter, (i.e., a composite excitation of the spin and charge
 degrees of freedom) with the larger energy scale.

\section*{ACKNOWLEDGMENTS}
%
 One of the author (H.O.) is grateful to
 Y. Okabe
 for helpful discussions.
 M.N. thanks J. Voit for the collaboration in the early stage of the
 present work. M.N. is partly supported by the Ministry of Education,
 Culture, Sports, Science and Technology of Japan through Grants-in-Aid
 No.\ 14740241.
 Main computations were performed using the facilities of
 Tokyo Metropolitan University,
 Yukawa Institute for Theoretical Physics, 
 and 
 the Supercomputer Center, Institute for Solid State Physics, University
 of Tokyo.
%

\end{document}